\documentclass[reprint,superscriptaddress,showpacs,amsmath,amssymb,
longbibliography,prl]{revtex4-1}

\usepackage{graphicx}
\usepackage{dcolumn}
\usepackage{bm}
\usepackage{color}
\usepackage{eufrak}
\usepackage{multirow}
\usepackage{pifont}

\newcommand{\un}[1]{\ensuremath{\,\text{#1}}}
\newcommand{\unit}[1]{\un{#1}}
\newcommand{\vb}{\ensuremath{V_{\text{b}}}}
\newcommand{\vbg}{\ensuremath{V_{\text{g}}}} 
\newcommand{\vg}{\vbg}

\newcommand{\Vg}{\ensuremath{V_{\mathrm g}}}

\newcommand{\vF}{\ensuremath{v_{\mathrm F}}}

\begin{document}

\title{Secondary electron interference from trigonal warping in clean 
carbon nanotubes}

\author{A. Dirnaichner}
\affiliation{Institute for Experimental and Applied Physics, 
University of Regensburg, 93040 Regensburg, Germany}
\affiliation{Institute for Theoretical Physics, University of Regensburg, 
93040 Regensburg, Germany}
\author{M. del Valle}
\affiliation{Institute for Theoretical Physics, University of Regensburg, 
93040 Regensburg, Germany}
\author{K. J. G. G\"otz}
\author{F. J. Schupp}
\author{N. Paradiso}
\affiliation{Institute for Experimental and Applied Physics, 
University of Regensburg, 93040 Regensburg, Germany}
\author{M. Grifoni}
\affiliation{Institute for Theoretical Physics, University of Regensburg, 
93040 Regensburg, Germany}
\author{Ch. Strunk}
\author{A. K. H\"uttel}
\email[]{andreas.huettel@ur.de}
\affiliation{Institute for Experimental and Applied Physics, 
University of Regensburg, 93040 Regensburg, Germany}

\date{\today}

\begin{abstract}
We investigate Fabry-Perot interference in an ultraclean carbon nanotube 
resonator. The conductance shows a clear superstructure superimposed onto 
conventional Fabry-Perot oscillations. A sliding average over the fast 
oscillations reveals a characteristic slow modulation of the conductance as a 
function of the gate voltage. We identify the origin of this secondary 
interference in intervalley and intravalley backscattering processes which 
involve wave vectors of different magnitude, reflecting the trigonal warping 
of the Dirac cones. As a consequence, the analysis of the secondary interference 
pattern allows us to estimate the chiral angle of the carbon nanotube.
\end{abstract}


\maketitle

Clean carbon nanotubes (CNTs) are an excellent material system to observe 
Fabry-Perot interference when highly transparent contacts suppress charging 
effects~\cite{Liang2001}. This is often the case in the hole regime of 
transport in CNTs~\cite{Grove-Rasmussen2007, Kamimura2009}. So far, experiments 
mostly concentrated on the effects of the linear, Dirac-like part of the 
CNT dispersion relation, resulting in simple Fabry-Perot (FP) 
interference~\cite{Liang2001, Kim2007, Man2006, Herrmann2007}. Its hallmark  
is an oscillatory behavior of the differential conductance $G(V_{\rm g},V_{\rm 
b})$ as a function of both gate voltage $V_{\rm g}$ and
bias voltage $V_{\rm b}$,  with frequency proportional to the CNT length 
\cite{Liang2001}. On top of this regular oscillation, \textit{slower} 
modulations are sometimes observed in experiments \cite{Liang2001, 
Kong2001, Man2006}. Such secondary interference has been attributed to disorder 
\cite{Kong2001, Romeo2011}, or to channel mixing at the CNT-contact interface 
\cite{Jiang2003}. It has been suggested that a slow modulation can also 
originate from intrinsic interference effects in chiral CNTs \cite{Yang2005}. 
In general, being related to a difference of accumulated phases,
secondary interference probes the nonlinearity of the CNT dispersion relation 
due to the trigonal warping, and in turn the chiral angle \cite{Jiang2003, 
Yang2005}.

In this Letter we report on the investigation of a peculiar secondary 
interference pattern in the hole regime  of an ultraclean CNT. Upon averaging 
over the fast primary FP-oscillations, the resulting average linear conductance 
$\bar{G}(V_{\rm g})$ shows a quasi-periodic slow modulation deep in the
hole regime. We combine detailed tight binding calculations and fundamental 
symmetry arguments to identify the origin of the slow modulation. Our analysis 
of the gate voltage dependence of $\bar{G}(V_{\rm g})$ allows us to estimate 
the CNT's chiral angle $\theta$. 

\begin{figure*}
\centering
\includegraphics[width=\textwidth]{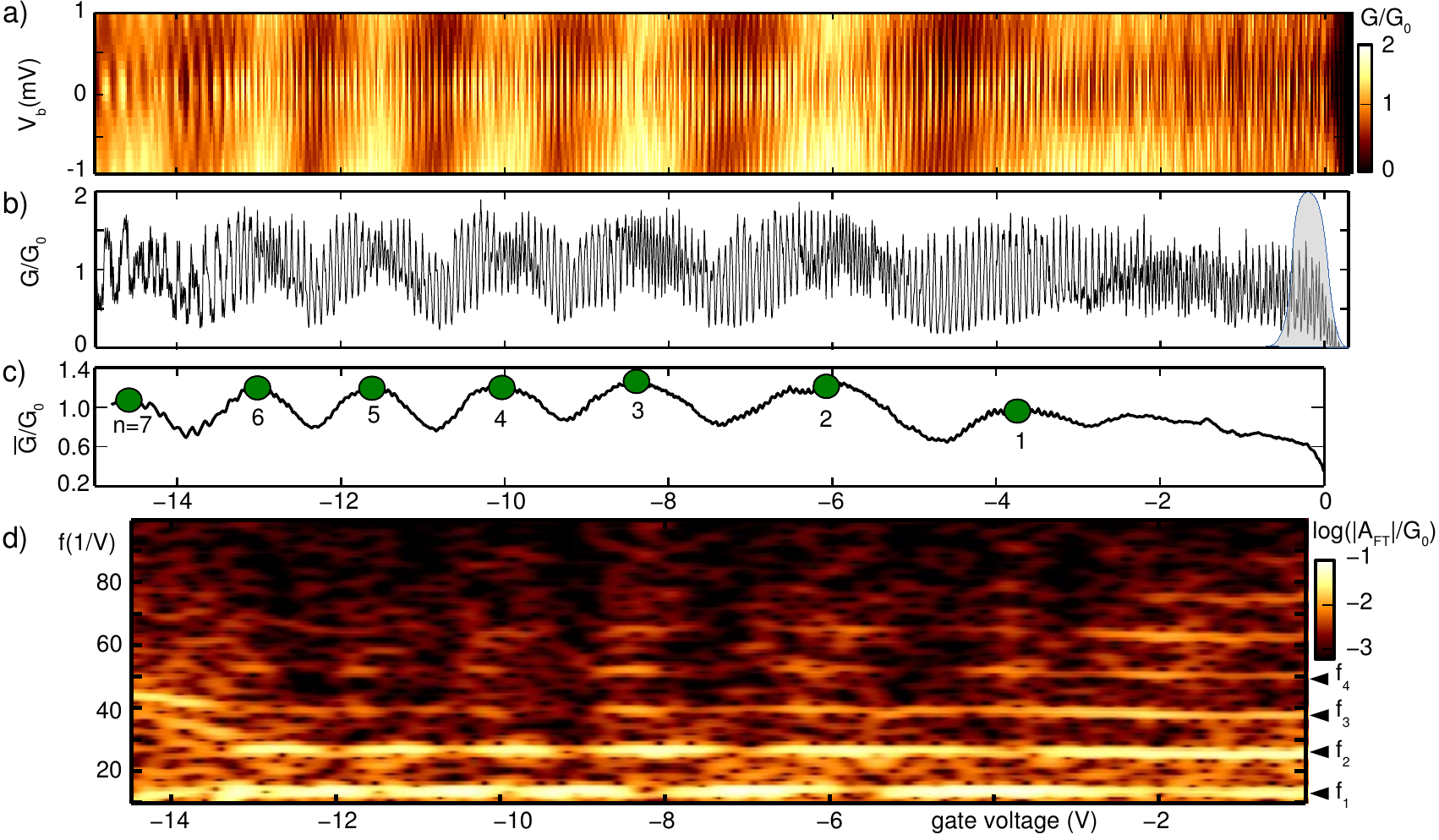}
\caption{(Color online)
(a) Differential conductance $G(\vg,\vb)$ of a clean carbon nanotube (CNT) 
device in the hole conduction regime, as a function of back gate voltage \vg\ 
and bias voltage \vb\ ($G_0=e^2/h$). 
(b) Zero bias conductance $G(\vg)$ extracted from (a). 
(c) Average conductance  $\bar G(\vg)$ obtained over a sliding $0.4\un{V}$-wide 
Gaussian window. A slow modulation is observed. The peak positions are marked 
with filled circles. The distance $\Delta V_{\mathrm{g},n} = V_{\mathrm {g},n} - 
V_{\mathrm {g},n+1}$ between the $n$-th and the $n-1$-th peak decreases with 
$n$. 
(d) Fourier transform of a sliding $0.4\un{V}$-wide window in the signal in (b) 
as a function of the gate voltage.
\label{fig:overview}} 
\end{figure*}   
We measure the differential conductance of a suspended CNT attached to 
$50\un{nm}$-thick Pt/Ti leads, separated by a $1.2$\unit{$\mu$m}-wide trench, 
at $T=15\unit{mK}$~\cite{supplement}. The fabrication process is optimized to 
produce defect free devices~\cite{Huttel2009}. Fig.~\ref{fig:overview}(a) 
displays the conductance $G(\vg,\vb)$ of the CNT device as function of gate 
voltage \vg\ and bias voltage \vb. On the electron conduction side ($\vg> 
0.35\un{V}$, see Suppl.~Mat.~\cite{supplement}), transport characteristics are
dominated by Coulomb blockade. On the hole side, owing to the high transparency 
of the barriers, the CNT behaves as an electronic 1D waveguide. An oscillatory 
large conductance, $0.2 \lesssim G/G_0 \lesssim 1.8$ ($G_0=e^2/h$), is observed 
for gate voltage values $-15\un{V} \lesssim \vg \lesssim 0\un{V}$. The electron 
wavevector is affected by both bias and gate voltage, leading to typical 
rhombic interference structures in the $G(\vg,\vb)$ diagram~\cite{Liang2001}.
A striking feature of our data is the slow modulation of the conductance 
pattern as a function of $\vg$, visible as a  series of darker and brighter 
intervals in Fig.~\ref{fig:overview}(a) alternating on a scale of ca. $2\un{V}$.

In Fig.~\ref{fig:overview}(b) we show the differential conductance trace 
$G(\vg)$ for $\vb = 0$. Primarily, we observe a fast oscillation of the
conductance at a frequency $f_1 = 12.8$~V$^{-1}$. This fundamental frequency is
directly related to the length of the cavity via $f_1\simeq\alpha e L /\pi\hbar 
\vF$~\cite{Liang2001}. For $\vF=8\cdot 10^{5}$\unit{m/s} 
\cite{Dresselhaus1996}, we obtain $L \simeq 1$\unit{$\mu$m} which is close to 
the width of the trench. From the period of the fast oscillation, $\Delta 
\Vg^{\mathrm {fast}}=1/f_1$ and the height $V_{\mathrm{c}}$ of the rhombic
pattern in Fig.~\ref{fig:overview}(a), we extract the gate voltage lever arm, 
$\alpha=V_{\mathrm c}/\Delta V^{\mathrm {fast}}_{\mathrm g}=0.0210\pm 
0.0007$ \cite{Liang2001}. On top of the fast oscillations the slow modulation 
is visible. Fig.~\ref{fig:overview}(c) shows the sliding average $\bar G(\vg)$ 
of the conductance as function of $\vg$. The peaks of the average conductance 
are labeled as $n=1\dots 6$ starting from the bandgap. The spacing of the peak 
positions $E_n=\alpha\Delta V_{\mathrm{g},n}$ decreases for more negative gate 
voltages $\vg$.

We perform a discrete Fourier transform (FT) over a Gaussian window (shaded 
gray in Fig.~\ref{fig:overview}(b)). The result is plotted in log-scale in 
Fig.~\ref{fig:overview}(d) as a function of frequency and window 
position. The FT shows regions in \vg\ with a dominant fundamental frequency 
component $f_1$ alternating with regions where the second harmonic with 
$f_2=2f_1$ prevails. These reflect the frequency doubling that is visible in 
certain ranges of $\vg$ in Fig.~\ref{fig:overview}(b). In these regions, the FT 
reveals also components from higher harmonic frequencies $f_n=nf_1$, appearing 
as horizontal lines in the FT plot. An analysis of the decay of the higher 
harmonic amplitudes in the Fourier transform yields an average length of the 
electronic path in the interferometer of $2.7$\unit{$\mu$m}. This length 
corresponds to the dwell time of an electron in the device and provides a lower 
bound on its phase coherence length~\cite{supplement}. As shown in 
Fig.~\ref{fig:overview}(d), the entire spectrum consists mainly of one 
fundamental frequency and its harmonics, i.e., no additional fundamental 
frequency occurs. Hence, we can conclude that there are no impurities that 
subdivide the CNT into a serial connection of multiple FP 
interferometers~\cite{Romeo2011}.

\begin{figure*}
\centering
\includegraphics[width=\textwidth]{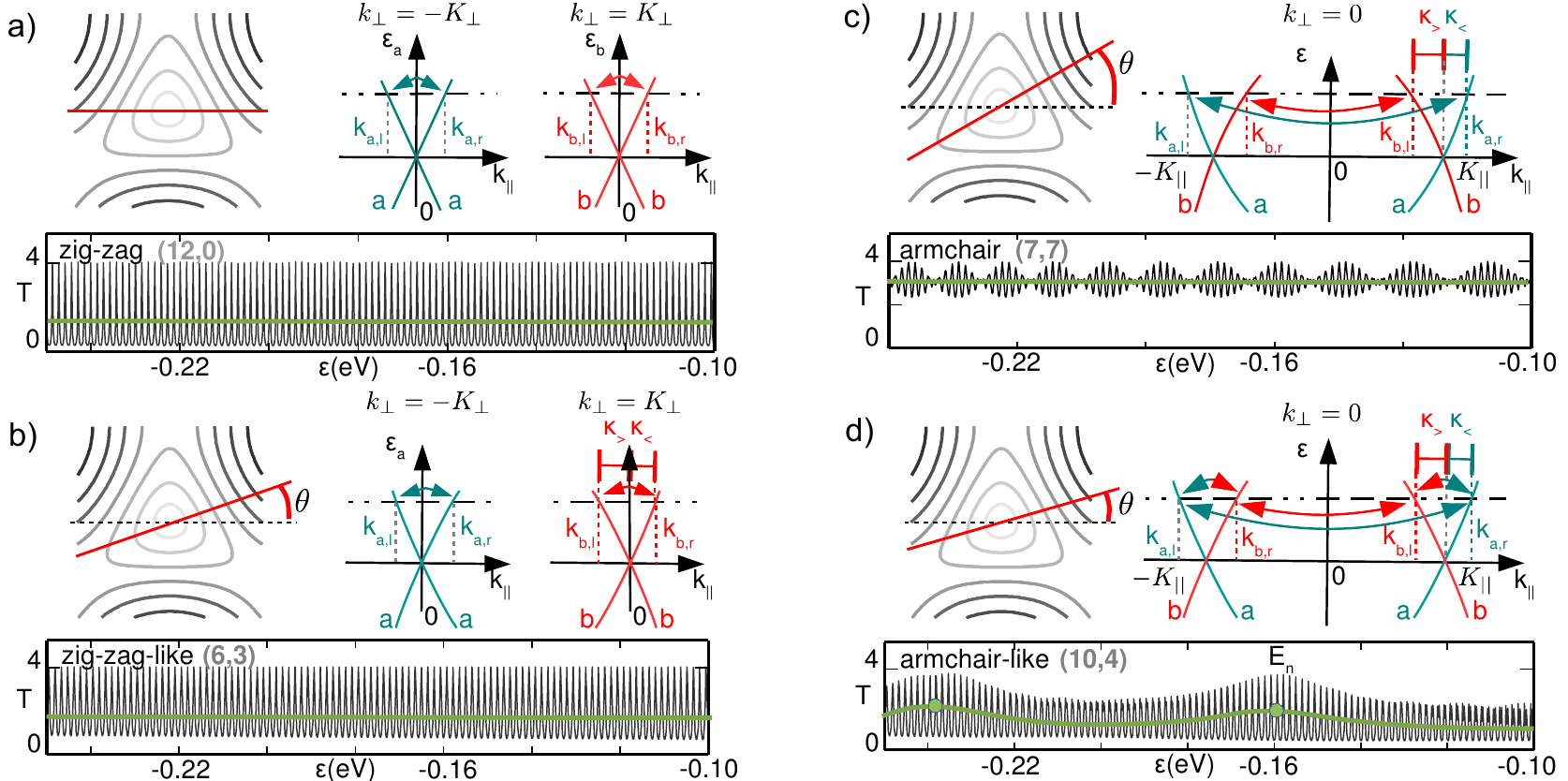}
\caption{(Color online)
Graphene dispersion relation $\varepsilon(\mathbf{k})$ (contour plots in the
top left panel of each subfigure) in the vicinity of a Dirac point and 
simplified lowest 1D subbands (line plots, top right) \cite{SOcomment}. The 
solid red line in the contour plot marks the direction of $k_{||}$. The chiral
angle $\theta$ is measured with respect to the direction of the \textit{zigzag} 
CNT (dashed line). The bottom panels show exemplary transmission patterns 
obtained by numerical tight-binding calculations. The green line represents the 
sliding average $\bar T$ of the transmission signal.
(a) \textit{Zigzag}. Dispersion relations at the two Dirac points $-K_\perp$ 
(green) and $K_\perp$ (red) are identical and symmetric with respect to the 
$k_{||}=0$ axis. The transmission curve of a (12,0) CNT shows a simple, single 
channel FP interference pattern.
(b) \textit{Zigzag-like}. Right and left moving branches within each valley 
exhibit different wave vectors $k_{j,r/l}$ at finite energy. No inter-valley 
scattering is possible in (a) and (b), see text. A single-channel-like
transmission pattern can be observed for the (6,3) CNT (bottom left).
(c) \textit{Armchair}. Parity symmetry forbids scattering between $a$ and $b$
branches. At finite energy, the two Kramers channels $a$ and $b$ have different 
wavevector associated to right- and left-moving states and a beat in the 
interference pattern is observed in the tight-binding calculation for the (7,7) 
CNT.
(d) \textit{Armchair-like}. In the armchair-like CNTs the parity symmetry is 
broken and inter-channel scattering is enabled (see text). A slow modulation of 
the transmission pattern can be observed in the average transmission $\bar T$ 
of a (10,4) CNT (bottom right).
\label{fig:model}} 
\end{figure*}

The main features observed in the experiment can be reproduced by a real-space
tight-binding calculation, a description that allows us to realistically 
include curvature effects and the spin-orbit interaction in the real-space 
Hamiltonian of our system~\cite{SOcomment, delValle2011}. The transport 
properties of the CNTs are obtained within the Landauer-B\"uttiker 
approach, using Green's function techniques, very well suited for transport 
calculations in the ballistic regime. This numerical approach can be applied to 
CNTs with arbitrary structure. In Fig.~\ref{fig:model}(a-d) our numerical 
results for the transmission of four different classes of CNTs are shown. 
Strikingly, the slow modulation pattern in the average transmission $\bar T$ is 
observed only for the CNT geometry in Fig.~\ref{fig:model}(d), where even in an 
idealized system the absence of certain symmetries (discussed below) allows 
interferometer channel mixing. As we are going to explain, the crucial 
geometrical property determining this secondary FP pattern is the chiral angle.

Carbon nanotubes can be grouped in four distinct classes~\cite{Lunde2005, 
Marganska2015, Laird2015}: \textit{armchair, armchair-like, zigzag} and 
\textit{zigzag-like}. The CNT chiral indices $(n,m)$ determine the class: If 
the ratio of $n-m$ to their greatest common divisor $d=\gcd(n,m)$ is a multiple 
of $3$, i.e., $\frac{n-m}{3\cdot d}\in\mathbb{Z}$, the CNT belongs to the chiral
\textit{armchair-like} class if $n\neq m$, and is an achiral \textit{armchair} 
CNT if $n = m$. Otherwise we are dealing with a \textit{zigzag-like} CNT unless 
$m=0$, which characterizes achiral \textit{zigzag} CNTs.

This classification reflects intrinsic differences in the CNT band structure, 
which are of crucial importance to the transport properties of these systems. 
In metallic \textit{zigzag} and \textit{zigzag-like} CNTs the $\pi$-bands 
cross at the Dirac points $\vec{K}=(k_{||}=0, k_\perp=+K_\perp)$, $\vec{K}' = 
-\vec{K}$ \cite{Lunde2005}. Here, $k_{||}$ and $k_{\perp}$ are the components 
of the wavevector parallel and perpendicular to the CNT axis. In particular, 
$k_{\perp}$ is proportional to the crystal angular momentum which
stems from the rotational ${\cal C}_d$ symmetry and is opposite in the two 
valleys. When considering reflections from the interfaces, this symmetry only
allows for {\it intravalley} backscattering. Thus each valley constitutes an 
independent transport channel, as depicted in Fig.~\ref{fig:model}(a,b). In 
this case, the FP oscillations are mainly described by the standard expression 
for the transmission~\cite{Datta1997},
\begin{align}\label{eq:transmission}
{\cal{T}}(\Vg)=\sum_{j=a,b}\frac{2|t_1|^2|t_2|^2}{1+|r_1|^2|r_2|^2-2|r_1||r_2|\cos[\phi_j(\Vg)]},
\end{align}
where $j$ labels the two independent channels, and the transmission and 
reflection amplitudes for the two confining barriers are given by $t_{1}$, 
$t_2$, and $r_{1}$, $r_2$, respectively. The phase accumulated upon one round 
trip is given by $\phi_j(\Vg)=\left( \left| 
k_{j,l}(\Vg)\right|+\left|k_{j,r}(\Vg)\right|\right) L$, and 
the wave 
vector of the right(left) moving electron $k_{j,r(l)}$ is linked to \Vg\ via 
the CNT dispersion relation $\varepsilon(k_{j,r(l)})=\alpha e\Vg$. In 
\textit{zigzag} and \textit{zigzag-like} CNTs, the accumulated phases are 
identical for the two channels since the dispersion in the two valleys is 
symmetric, i.e., $k_{a,r}=\left|k_{b,l}\right|$, and 
$k_{b,r}=\left|k_{a,l}\right|$. According to 
Eq.~(\ref{eq:transmission}), one single FP oscillation occurs when 
$\phi_j=2\pi$. Consequently, the tight-binding model calculations of a $(12,0)$ 
CNT in Fig.~\ref{fig:model}(a) and of a $(6,3)$ CNT in (b) show featureless 
single-channel interference patterns with a fundamental frequency $f_1$.

On the other hand, in \textit{armchair} and \textit{armchair-like} CNTs the bands 
cross at the Dirac points $\vec{K}=(K_{||}, 0)$ and $\vec{K}'=-\vec{K}$, see 
Fig.~\ref{fig:model}(c,d). Two valleys are formed, which are symmetric with 
respect to the $k_{||}=0$ axis and both characterized by zero crystal angular 
momentum~\cite{Lunde2005}. Intervalley backscattering is now possible and the 
angular momentum quantum numbers do not provide a mean to distinguish the 
transport channels.

However, \textit{armchair} CNTs are invariant under the parity operation 
\cite{Lunde2005}, which enables us to identify now two other independent 
transport channels, $a$ and $b$. These parity channels are such that within one 
pair, backscattering connects a left-mover or right-mover in the $K$ valley to 
its time-reversal partner in the 
$K'$-valley, see Fig.~\ref{fig:model}(c). Eq.~(\ref{eq:transmission}) still 
describes the FP oscillations but, in contrast to the \textit{zigzag-like} 
class, the two channels accumulate different phases, $\phi_a = 2k_{a,r}L \neq 
\phi_b = 2k_{b,l}L$, owing to the trigonal warping. In the interference pattern 
we thus expect a \textit{beat} with a constant average transmission. This 
expectation is confirmed by our tight-binding transport calculations for a 
$(7,7)$ CNT, see Fig.~\ref{fig:model}(c).

In \textit{armchair-like} CNTs, the parity symmetry is absent and hence 
backscattering from branch $a$ to branch $b$ in the same valley is also 
possible. The interference pattern displays secondary interference with slow 
oscillations of the average transmission. The occurrence of the slow modulation 
can be understood from the mode-mixing within a simplified model, see Suppl. 
Mat.~\cite{supplement}. This observation is confirmed by the tight-binding 
modeling of a $(10,4)$ CNT in Fig.~\ref{fig:model}(d). Our calculation clearly 
demonstrates that valley mixing effects can occur also in clean CNTs 
\cite{Schmid2015}, and cannot be taken as an indicator of disorder.

In a realistic experiment, the coupling between the CNT and the metallic 
contacts differs between CNT top and bottom parts, and depends on the 
fabrications details. In the Supplement we have investigated the effects of an 
extrinsic top/bottom symmetry breaking at the contacts in \textit{zigzag-like} 
CNTs. It induces a breaking of the rotational ${\cal C}_d$ symmetry, and hence 
allows for transport channel mixing. Tight-binding calculations confirm that 
then a slow modulation of $\bar G$ analogous to the \textit{armchair-like} case 
emerges \cite{supplement}. 

For a quantitative analysis we extract the peak positions $E_{n} = \alpha 
V_{\mathrm g,n}$ of the slow modulation of the average conductance $\bar G$ 
[green dots in Fig.~\ref{fig:overview}(c)], and compare these values to 
theoretical predictions. A simple model (see Suppl. Mat. \cite{supplement}) 
shows that the slow modulation is governed by the phase difference between 
Kramers channels $\Delta \phi^\theta(E) = 2\left(\kappa_>^\theta - 
\kappa_<^\theta\right)L$. Here, the $\kappa_>^\theta \ge \kappa_<^\theta \ge 0$ 
are the longitudinal wave vectors measured from the same Dirac point. In an 
\textit{armchair-like} CNT with chiral angle $\theta$, $k^{\theta}_{a,l} = 
-K_{||}-\kappa^{\theta}_<$, 
$k^{\theta}_{b,r} = -K_{||} + \kappa^{\theta}_>$, $k^{\theta}_{b,l} = 
K_{||}-\kappa^{\theta}_>$ and $k^{\theta}_{a,r} = K_{||} + 
\kappa^{\theta}_<$, see Fig.~\ref{fig:model}(d).
For a \textit{zigzag-like} CNT, the $\kappa^{\theta}_{>/<}$ are in analogy 
given by $\kappa^{\theta}_>=k_{a,r}^\theta = |k_{b,l}^\theta|$ and 
$\kappa^\theta_<=|k_{a,l}^\theta|=k_{b,r}^\theta$, see 
Fig.~\ref{fig:model}(b). In either case, a peak occurs when
\begin{align}\label{eq:phase-rel}
\Delta \phi^{\theta}(E)&=2\pi n.
\end{align}
This result is validated by tight-binding calculations~\cite{supplement}. 

\begin{figure}
\centering
\includegraphics[width=\columnwidth]{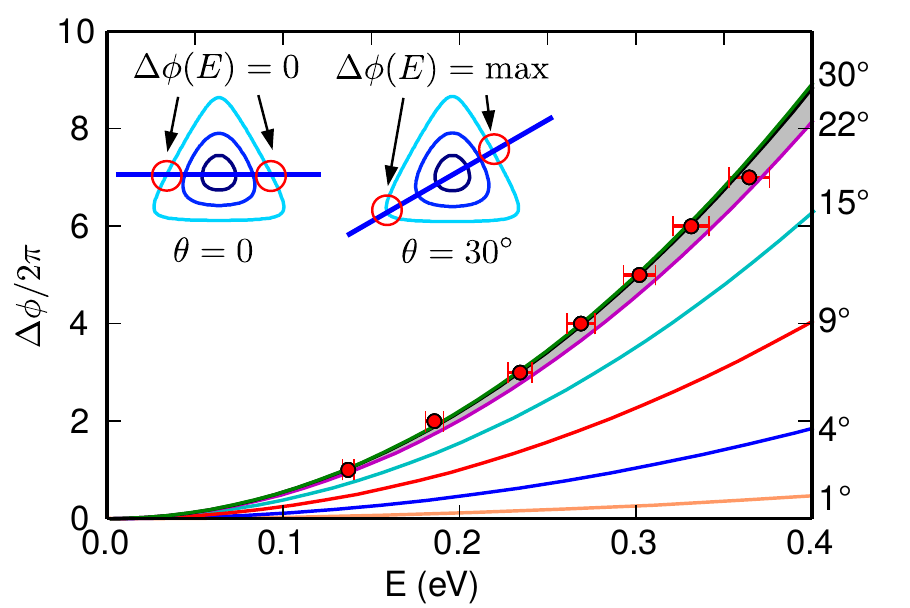}
\caption{(Color online)
Computed phase differences $\Delta \phi^\theta$ between modes as a function of 
energy measured from the Dirac point for different chiral angles $\theta$. The 
phase difference is a monotonically increasing function of $\theta$ starting 
from the \textit{zigzag} CNT with $\theta=0$ (inset, left) to the 
\textit{armchair} CNT with
$\theta=30^\circ$ (inset, right). The filled circles are obtained using the 
experimental positions $E_n+\Delta E_{\mathrm {gap}}$ of the slow modulation 
and requiring $\Delta \phi/2\pi = n$. The error bars indicate the uncertainty 
in $\alpha$ and in $\Delta E_{\mathrm{gap}}$ (see text). Acceptable fits are 
obtained by chiral angles in the range $22^{\circ} \leq \theta < 
30^{\circ}$~\cite{supplement}, as indicated by the gray shaded area. 
\label{fig:dkvsE}} 
\end{figure}   
The phase difference $\Delta \phi^{\theta}(E)$ is computed
numerically from the tight-binding dispersion relation
$\varepsilon^{\theta}(k^\theta_{j,i})$~\cite{Saito1998}. It
is shown for different chiral angles $\theta$ in Fig.~\ref{fig:dkvsE}. The 
slope of $\Delta \phi^{\theta}(E)$ is monotonically increasing with $\theta$ 
and is zero for the \textit{zigzag} case (Fig.~\ref{fig:dkvsE}, left inset) 
and maximal for the \textit{armchair} case (right inset). In the model 
calculation the energy is measured from the Dirac point. In the experiment, 
however, the center of the gap is located at $\vg = 0.31\un{V}$. Hence, to 
check whether the experimental peak positions are determined by 
Eq.~(\ref{eq:phase-rel}), one has to account for an energy shift $\Delta 
E_{\mathrm{gap}} = \int_{0\un{V}}^{0.31\un{V}} \alpha_{\mathrm{gap}}(\vg) 
\mathrm{d}\vg$, where $\alpha_{\mathrm {gap}}(\vg)$ is the lever arm in the gap 
region. $\alpha_{\mathrm {gap}}(\vg)$ increases in the vicinity of the bandgap 
starting at $\vg=0.15\un{V}$ until it reaches $0.68\pm 0.03$ within the 
bandgap~\footnote{ The origin of the $60\pm 5\un{meV}$ bandgap in our CNT 
can not be explained from the CNT curvature. A curvature induced gap of 
$<20\un{meV}$ is estimated for a CNT with $\theta>22^{\circ}$. Note that for our 
analysis, the nature of the small bandgap is not crucial. We focus on energies
$\varepsilon$ larger than $90\un{meV}$ where the effect of the finite bandgap 
on the dispersion is negligible.}. The dots in Fig.~\ref{fig:dkvsE} are thus 
given by the coordinates $(E_n+\Delta E_{\mathrm {gap}},2\pi n)$ and are 
compared to $\Delta \phi^{\theta}(E)$. The chiral angle $\theta$ can thus be 
used as a fit parameter. The error bars indicate the experimental uncertainty 
for $\Delta E_{\mathrm {gap}}$, which we are only able to restrict to 
a range $55\un{meV}<\Delta E_{\mathrm {gap}}<60\un{meV}$, and $\alpha$ 
\cite{supplement}. The fit provides an estimation of $22^{\circ} \leq \theta < 
30^{\circ}$ for the chiral angle, see Fig.~\ref{fig:dkvsE} (gray shaded area).

At high energies the chiral angle $\theta$ and the trigonal warping of the 
graphene dispersion relation alone determine slope and curvature of the 1D 
subbands and thereby the accumulated phase difference between the Kramers 
channels \cite{supplement}. In a realistic experiment there is likely an 
extrinsic symmetry breaking at the contacts. Thus, channel mixing is expected 
for both \textit{zigzag-like} and \textit{armchair-like} CNTs, and in either 
case allows for evaluation of the chiral angle when several periods of the slow 
modulation are recorded.

In conclusion, the secondary Fabry-Perot interference
provides a robust tool to estimate the chiral angle, a key characteristic which 
is crucial for understanding carbon nanotube properties such as the spin-orbit 
coupling~\cite{Kuemmeth2008, Steele2013} or the KK' mixing 
\cite{Marganska2015}. In contrast to other methods as, e.g., Raman spectroscopy 
or scanning probe microscopy, which are difficult to combine with transport 
spectroscopy, our analysis can be easily integrated with measurements in the 
few-electron or in the Kondo regime.

The authors acknowledge financial support by the Deutsche Forschungsgemeinschaft
(Emmy Noether grant Hu 1808/1, GRK 1570, SFB 689).

\end{document}